# High-order harmonic generation in 2D Transition Metal Disulphides

J. M. Iglesias[1,@], E.Pascual[1], M. J. Martín[1], and R. Rengel[1]

[1] Department of Applied Physics, University of Salamanca.
Plaza de la Merced S/N (Edificio Trilingüe), 37008, Salamanca (Spain)
[@] josem88@usal.es

## ABSTRACT

In this paper we explore the capabilities of $MoS_2$ and $WS_2$ 2D monolayers to produce radiation in the terahertz range by generation of high-order harmonics. This phenomenon, which is a result of the non-linear response of the electronic carrier population to the applied electric field, is studied by using a particle ensemble stochastic simulation approach based on the Monte Carlo method. The power of the produced harmonic signals is studied against the electric field amplitude, the external temperature, and the frequency of the excitation. Additionally, the stochastic nature of the simulation tool enables to discern the purely discrete harmonic signal from the background spectral noise that comes from the intrinsic carrier velocity fluctuations in the diffusive regime, permitting to set bandwidth thresholds for harmonic extraction. It was found that both TMDs showed similar thresholds bandwidths when compared to III-V semiconductor at low temperatures, while $WS_2$ would be by far a better choice, over $MoS_2$, for exploitation of the $7^{th}$ and $9^{th}$ harmonic generation.

Finding reliable sources of electromagnetic radiation between the microwave and the infrared spectral domains remains a challenge in the solid-state electronics field[12,3]. Because of this, there exists legitimate interest in the discovery and exploitation of new materials and device contepts that could fulfill the growing demands for promising applications, which include imaging, short-range high data rate communication, sensing, or spectroscopy. One approach for the development of THz applications is the investigation of new materials with the potential to outperform the current technology. In this context, the demonstration of the stability of two-dimensional crystals that came with the discovery of graphene[4] opened op new promising research lines for the future of electronics, including high-frequency applications[5]. Graphene has got plenty of attention from the research community due to its excellent electronic properties –specially, its elevated carrier mobility[6]–. However, being a semi-metal, its in-practice zero band gap makes it unsuitable for applications where effective device switching off is needed, as band-to-band tunneling[7,8] and interband carrier-carrier scattering[8] phenomena ensure the presence of mobile charges, and thus the difficulty to completely turn off a graphene channel. In this sense, monolayer transition metal dichalcogenides (TMDs)[9,10,11], albeit showing notably much lower electron and hole mobilities[12,13,14,15] in comparison with carbon monolayers, their band gaps, which lie above 1 eV[11,16,17,18], places them in suitable positions for applications in the next generation of solid state electronic devices[19,20,21,22].

In this paper the feasibility of THz emission from TMDs via harmonic generation of readily available sources for alternate current is studied. The collective non-linear velocity response of charge carriers (drift velocity) to an alternating (AC) electric field described as $E(t) = E_f \cos(2\pi f t)$, where $E_f$ is the amplitude and $f$ the oscillation frequency, contains harmonics of such fundamental excitation, and therefore radiation in these frequencies would be expected, as depicted in FIG. 1. In this analysis we will make use of an in-house semi-classical Ensemble Monte Carlo (EMC) simulator[23,24] for microscopic electronic transport, which provides access to useful information like individual scattering activity, or quantities derived from fluctuations thanks to its stochastic approach. As for the physical model, the bandstructure is described by means of parabolic $\varepsilon$-**k** dispersion relations for the valleys situated at K and Q points[16] of the first Brillouin zone. The consideration of the upper Q valleys is mandatory

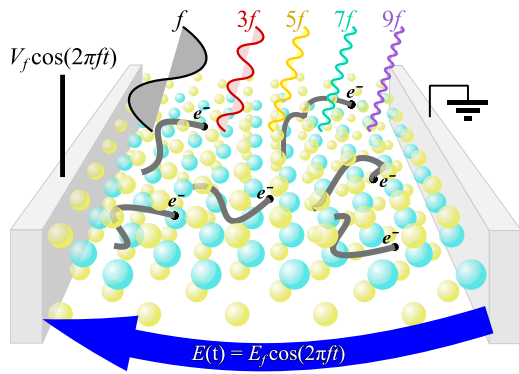

**FIG. 1** Illustration of harmonic radiation with the application of an alternating electrif field to a TMD monolayer.

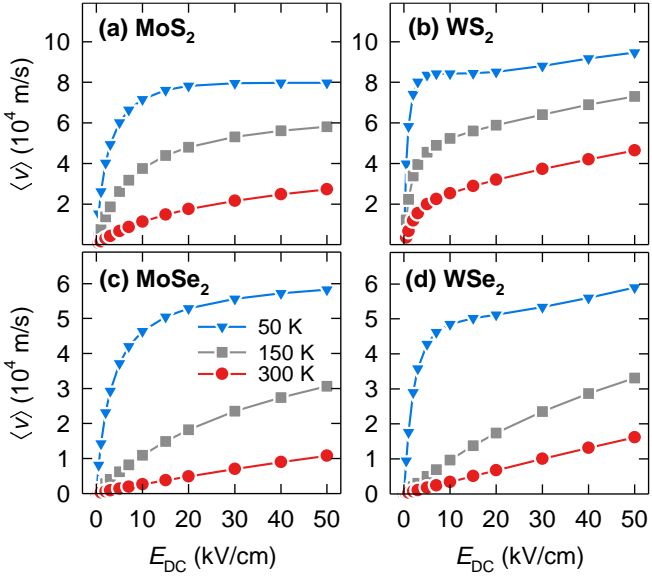

**FIG. 2** Electron drift velocity as a function of the in-plane electric field for (a) MoS$_2$, (b) WS$_2$, (c) MoSe$_2$, and (d) WSe$_2$ at $T$ = 50, 150, and 300 K

for a correct description of electronic transport, as relatively strong in-plane electric fields will be reached in order to maximize the generation of harmonics, at it will be discussed later. Intrinsic (phonon) scattering phenomena is described by means of the deformation potential approximation of aggregated phonon modes[16,25] for possible transitions in the conduction band (Q and K intravalley, K ↔ K′, and K ↔ Q$_n$, and Q$_n$ ↔ Q$_{m \neq n}$). Further details of the model can accessed in the supplementary information.

In order to obtain a noticeable generation of high-order harmonics, it is necessary firstly to choose a material where the drift velocity-electric field dependence is prominently non-linear. The most extreme example of such non-linearities are III-V semiconductors, where hot electron transference between high and low mobility valleys leads to the Gunn effect[26]. Although in TMDs such negative differential resistance phenomenon is not present, harmonic generation still happens if there is a strong transition from the linear to saturation drift velocity regimes, as we have already observed in the case of graphene[27]. In FIG. 2 we show the ensemble electron drift velocity against the in-plane electric field for MoS$_2$, WS$_2$, MoSe$_2$, and WSe$_2$ at various temperatures. The transition metal diselenides are expected to show a much tempered harmonic generation due to a softer transition between linear behaviour (under small applied electric fields) and saturation velocity regime (for larger electric fields), and so we will focus this study to transition metal disulphides. Note that this work is restricted only to transport of electrons and consequently, the results must be interpreted in terms of n-type performance.

Let us now present the instantaneous ensemble velocity, ⟨$v$⟩ as a function of time when an alternate (AC) electric field is applied as depicted in FIG. 3 (a) and (d) as obtained by averaging the recorded instantaneous ensemble velocity at each instant within the period relative to the excitation frequency. The velocity response when $E_f$ is the smallest (5 kV/cm) retains most of the cosine-like waveform of the excitation electric field, so in these conditions, harmonics would be expected to be generated with rather limited power. Carrier inertia manifests from the dephasing of the velocity response and the electric field. The distortion of the waveform becomes progressively more evident $E_f$ rises. At the highest amplitudes the velocity response features strong flattening due to the velocity saturation observed in FIG. 2. This occurs when the electric field is close to the maximum absolute value (i.e., at $\theta = 0$, and $\theta = T_f/2$) being the time span larger for MoS$_2$ than for WS$_2$ under these operating conditions. The observed cycle-averaged scattering times range between 20 fs and 750 fs approximately, depending on the temperature and the material. The period of the applied signal is 5 ps (frequency 200 GHz), which indicates that the materials under study can reasonably respond (from a microscopic point of view) to the signal variations even under the coolest conditions. In these materials, the limitation of the maximum velocity is mainly determined by the occupation of the upper Q valleys, where the effective mass is larger, and the subsequent addition of further scattering mechanisms related to these valleys contributes to the reduction of the momentum relaxation time. In both materials at strong amplitudes a bump appears as the magnitude of the drift velocity rises, resembling transient velocity overshoots[28] that take place after a sudden field application takes place from equilibrium. This kind of transient responses have been found to be a source of THz radiation in III-V semiconductors when subject to ultrashort high field pulses[29], and so, these overshoots are expected to play a similar role in this study.

The numerical tool that must be used to evaluate the contribution of all the abovementioned features to harmonic generation is the calculation of the Fourier coefficients, $v_m$:

$$v_m = \frac{1}{T_f} \int_0^{T_f} \langle v(\theta) \rangle \exp(-2\pi i m f \theta)\, d\theta, \quad (1)$$

where the angular braces ⟨…⟩ mean averaging over the particle ensemble, and $m$ is a positive integer representing the harmonic order. The intensity of the $m^{th}$ harmonic is characterized by $|v_m|^2$. In FIG. 3 (b) and (e) we show the 1$^{st}$ to 9$^{th}$ odd harmonic coefficient squares plotted against $E_f$ in order to evaluate the power emitted at each of these frequencies under the same simulation parameters of FIG. 3 (a) and (d). As expected from the observed distortion in the drift velocity waveforms, the intensity of harmonics grows as the electric field amplitude is increased. For the 5$^{th}$, 7$^{th}$, and 9$^{th}$ harmonic orders, the variation in intensity is almost linear, ranging along up to five orders of magnitude, while for the 3$^{rd}$ harmonic most of the variation is limited to the lowest fields ($E_f < 10$ kV/cm), although in the particular case of WS$_2$, an inverse relation between $|v_3|^2$ and the electric field is noticeable above 20 kV/cm. Thus, exploitation of high-order harmonic generation will imply the application of rather strong electric fields amplitudes.

A further concern for a maximization of the generated power/intensity is the choice of the fundamental (excitation) frequency. In FIG. 3 (c) and (f) we show the Fourier coefficient squares versus frequency of the applied AC field in both monolayer TMDs. Fundamental and 3$^{rd}$ harmonic intensities show a very weak frequency dependence, while for

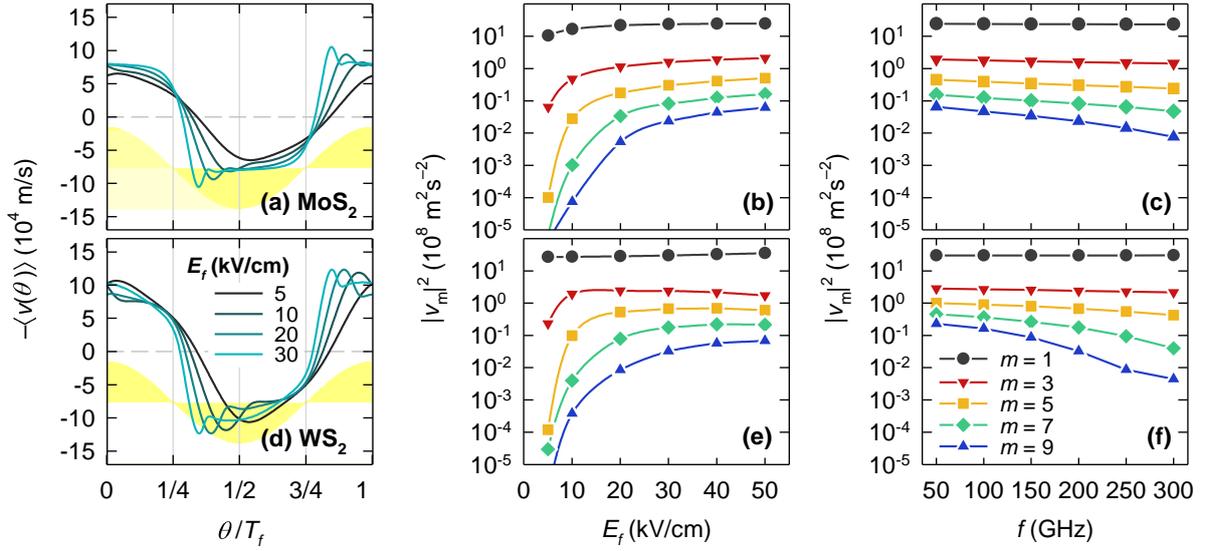

**FIG. 3** (a) and (d): Instantaneous ensemble drift velocity in (a) MoS$_2$ and (d) WS$_2$ as a function of the instant within the period of an alternating in-plane electric field of $f$ = 200 GHz and various amplitudes at $T$ = 50 K; the shaded region represents the electric field waveform. (b) and (e): Squares of the Fourier coefficients for the 1$^{st}$, 3$^{rd}$, 5$^{th}$, 7$^{th}$, and 9$^{th}$ harmonics as a function of $E_f$ considering an excitation frequency $f$ = 200GHz in (b) MoS$_2$ and (e) WS$_2$. (c) and (f): same as (b) and (e) as a function of the frequency with $E_f$ = 30 kV/cm and $T$ = 50 K in (c) MoS$_2$ and (f) WS$_2$.

the 5$^{th}$ to 9$^{th}$ harmonics, there is a noticeable loss of intensity as frequency is tuned up. The ideal case scenario for harmonic generation would be at a relatively low frequency, so that the electric field and the scattering mechanisms opposing forces allow the system to reach a quasi-equilibrium state and the instantaneous drift velocity corresponds to that of the instantanous electric field at stationary conditions, $\langle v(\theta) \rangle = \langle v \rangle|_{E=E(\theta)}$. The loss of harmonic generation efficiency at the highest sampled frequencies comes as a result of a combination of a comparatively slow drag of the electric field, and a decreasing effectivity of the scattering mechanisms to relax carrier momentum, leading to a more linear velocity response. For the forthcoming analysis we will focus in the 200 GHz excitation AC field, for being a readily available frequency in existing sources, and for being high enough to reach the THz range via harmonic generation.

The production and detection of harmonic radiation is not just reduced to the generation of harmonics with a particular intensity. Likewise in many other electronic applications related to electron transport, noise becomes a problem as it concerns signal detection. In order to accomplish a useful signal extraction, it must be of stronger intensity in comparison with the background noise at that particular frequency. Monte Carlo simulations for electronic transport are the ideal tool to successfully assess noise phenomena, as carrier fluctuations are considered in a straightforward and direct manner[30]. In transient simulation such as the ones at stake, the numerical tool to obtain the spectrum of electron velocity fluctuations, is the two-time correlation function, obtained from the instantaneous velocity history of each particle of the ensemble. It reads[31]:

(2) $\quad C_{\delta v \delta v}(\theta, s) = \langle v(\theta)v(\theta + s) \rangle - \langle v(\theta) \rangle \langle v(\theta + s) \rangle,$

where $s$ is the correlation time, equal to $t - t'$. Since the velocity response is periodic, we can consider $t$ as a time within the interval $[0, T_f]$, so $t = \theta + l\, T_f$, $l$ being an positive integer, and consequently, $\langle v(t) \rangle = \langle v(t + l\, T_f) \rangle = \langle v(\theta) \rangle$. Although we will not make further comments about the two-time correlation function, more information can be found in the supplementary material. The spectral density of velocity fluctuations, $\overline{S_{\delta v}}(\nu)$, is expressed by time-averaging the Fourier transform of the previously obtained correlation function over the AC field period as:

(3) $\quad \overline{S_{\delta v}}(\nu) = \frac{t}{T_f} \int_0^{T_f} \int_{-\infty}^{+\infty} C_{\delta v \delta v}(\theta, s) \exp(2\pi i \nu s)\, ds\, d\theta$

Now, our interest is set onto the comparison of the radiative power originated from the background noise related to random velocity fluctuations, and the power generated due to the collective response to the excitation electric field. The former intensity is expressed in terms of the Fourier coefficients as $2NT_f |v_m|^2$, where $NT_f$ is the exposure time under consideration, with $N$ being the number of cycles that the sensor would be exposed to the aforementioned radiation. Finally, the final spectral density, $S_v^T$ is the sum of the fluctuation and harmonic contributions:

(4) $\quad S_v^T(\nu) = \overline{S_{\delta v}}(\nu) + \sum 2NT_f |v_m|^2 \delta(\nu - \nu_m).$

Another procedure which allows obtaining the fluctuation and harmonic response contribution altogether is the finite Fourier transform of the instantaneous velocity from a representative subset of particles from the ensemble[32,31]:

(5) $\quad S_v^T(\nu) = 2NT_f \langle g(\nu)g^*(\nu) \rangle,$

where $g(\nu)$ is the individual particle Fourier coefficient given by:

(6) $\quad g(\nu) = \frac{1}{NT_f} \int_0^{NT_f} v(t) \exp(-2\pi i \nu t) dt\,.$

FIG. 4 shows the spectral density of the velocity response obtained by the two methods corresponding to equations (4) and (6) in both compounds at moderate and high AC electric field amplitudes. First, we note that the numerical correspondence proves the equivalence of both methods in the spectral description of the velocity response. The velocity fluctuation contribution varies qualitatively with $E_f$. At small fields – FIG. 4 (a) and (b)–, the background noise spectrums follow a lorentzian shape, with over an order of magnitude loss of spectral density within the sampled frequencies. Such a power reduction of the spectral fluctuation contribution would allow observing high order harmonics, which, on the other hand, in this low-field scenario are almost non-existent as seen before from the small values of the Fourier coefficients. At higher fields –FIG. 4 (c) and (d)–, the spectral density due to fluctuations flattens over the sampled frequencies. Although scattering probabilities, and thus momentum relaxation is lower in $WS_2$ compared to $MoS_2$, $\overline{S_{\delta v}}(\nu)$ is larger in the W compound in the studied frequency range. This is a consequence of larger single-particle velocities, which relate to the smaller K and Q valleys effective masses[16,24], and leads to widening the range for velocity fluctuations. At this AC electric field amplitude the harmonic contributions up the 9$^{th}$ order are well above the background noise.

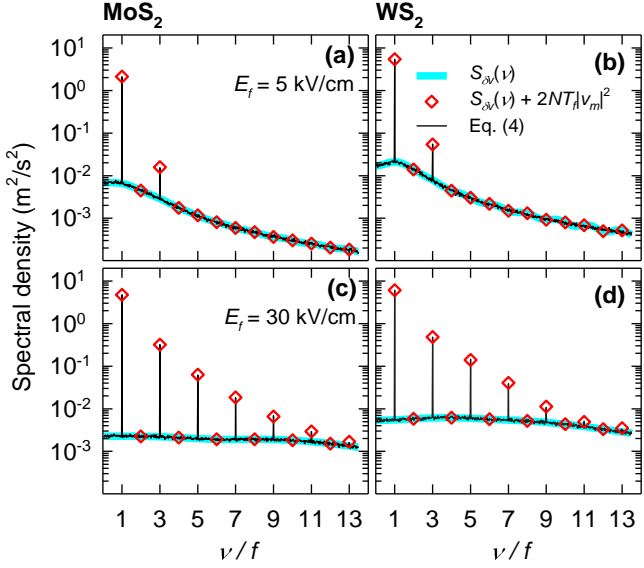

**FIG. 4** Spectral density of velocity response as a function the frequency normalized to that of the AC field, with $f = 200$ GHz and amplitudes of (a) and (b): 5 kV/cm, and (c) and (d): 30 kV/cm. Left panels (a) and (c) correspond to $MoS_2$ and (b) and (d) to $WS_2$ at $T = 50$ K. The number of cycles is $N = 200$.

A better evaluation of the high-ordear harmonic generation and its detection above the background noise can be done by means of the threshold frequency bandwidth[32]. This quantity can be defined as the bandwidth that a detector must have so that the regular (harmonic) and noise contributions are equal, and below which, the regular contribution is not detectable. It can be calculated in an approximated way as:

$$\Delta \nu_{\text{th},m} = 2|v_m|^2/\overline{S_{\delta v}}(\nu_m). \qquad (7)$$

In FIG. 5 we present the threshold frequencies obtained for the 3$^{rd}$, 5$^{th}$, 7$^{th}$ and 9$^{th}$ harmonics for both materials. In all the cases there is a monotonic trend in $MoS_2$ to increase the bandwidth threshold with $E_f$, while in $WS_2$ an optimum value is reached within the sampled range. In $MoS_2$ the maximum threshold badwidths are therefore obtained for the maximum $E_f$ of 50 kV/cm, being around 260, 65, 23 and 8 GHz for each odd harmonic order from the 3$^{rd}$ to the 9$^{th}$ respectively. In $WS_2$ under these conditions, this quantity lies in the same order of magnitude, being 90, 21, 8 and 2 GHz respectively. We note that these values are are comparable with those obtained with bulk III-V semiconductors, according to alike simulation scheme[32], and under similar low temperature conditions, with the exception of InP, that shows show peak values that clearly outperform the TMDs under study.

Finally, we explore the harmonic generation and extraction dependence with the temperature. In FIG. 5 we plot the frequency threshold as a function of the temperature for a fixed excitation of $E_f = 30$ kV/cm. By increasing the temperature, the resulting radiation spectra requires the bandwidth to be substantially reduced in order to detect the regular signal over the background noise. Such reduction in the regular to noise contribution ratio is a result of a drop in the harmonic generation efficiency. This can be microscopically explained by two factors that are direct consequences of increasing the temperature. First, the overall electron-phonon scattering activity is enhanced due to the

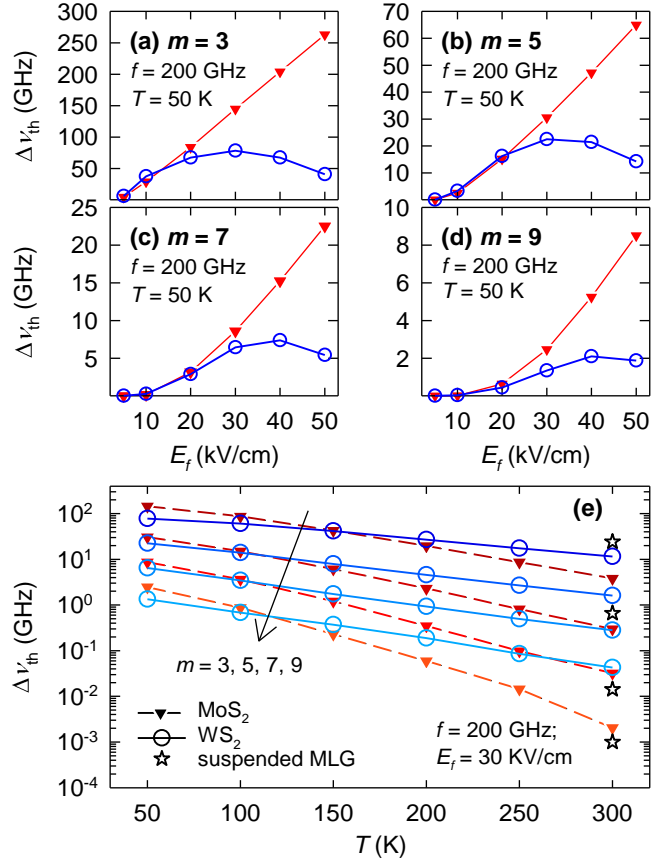

**FIG. 5** (a) to (d): Threshold bandwidths for the (a) 3$^{th}$, (b) 5$^{th}$, (c) 7$^{th}$ and (d) 9$^{th}$ harmonics as a function of $E_f$ in $MoS_2$ and $WS_2$ at $T = 50$ K. (d): Temperature dependence of the threshold bandwidths temperature dependence for the 3$^{rd}$, 5$^{th}$, 7$^{th}$ and 9$^{th}$ harmonic generation with an AC excitation electric field of 30 kV/cm in $MoS_2$ and $WS_2$. Stars represent the bandwidth thresholds for graphene under an excitation frequency of $f = 300$ GHz as obtained from ref.[27].

larger thermal phonon population, which reduces the mean free path and the mobility (see FIG. 2). The second factor is the attenuation, as temperature increases, of the dependence on electric field of the Q valley carrier occupation, as hotter carrier distributions involve higher occupations of upper valleys even at zero-field. This later factor results in the dissipation of the square-like shape of the collective velocity response waveforms, as Q↔K and Q↔Q scattering does not occur as a sudden onset when reaching high fields, but in a more gradual manner along each cycle. Graphical information supporting these claims can be found in the supplementary material. Also, as temperature is raised, the spectral density of velocity fluctuations is reduced in a similar way for all frequencies, but to a much lesser extent in comparison to the regular harmonic intensities. At ambient temperature, bandwidth thresholds can be compared to those obtained with graphene under an AC field of 10 kV/cm and frequency of 300 GHz; graphene outperforms both TMDs at the 3$^{rd}$ while WS$_2$ allows for noticeable broader bandwidths from the 5$^{th}$ harmonic, and MoS$_2$ shows similar values for the highest orders. However, it must be noted that these simulations are not carried out in the same conditions regarding the amplitude and frequency of the AC electric field: under the same operating conditions graphene would probably outperform both TMD materials.

In summary, we have made use of an in-house Ensemble Monte Carlo simulator to study the viability of production of THz radiation via high-order harmonic generation in bulk monolayer MoS$_2$ and WS$_2$. Collective velocity responses at low temperatures showed a velocity overshot and the approximation to a square-like waveform when exposed to a large enough AC excitation field intensities, which are necessary for optimal harmonic generation. Regular harmonic and fluctuation (noise) contributions to the spectral densities were identified using the two-time correlation function and the Fourier coefficients of the velocity response. In MoS$_2$ the threshold bandwith for the harmonics detection shows a monotonic trend with the electric field, while for WS$_2$ optimal AC field amplitudes can be spotted at fields below 50 kV/cm. MoS$_2$ and WS$_2$ showed a comparable feasibility for THz harmonic generation when compared to III-V semiconductors at low temperature, and also to freestanding graphene at room temperature.

## Supplementary material

Supplementary material includes details of the EMC model, and basic carrier transport data, along with some supporting results to the manuscript.

## Data availability

The data that support the findings of this study are available from the corresponding author upon reasonable request. The details of the methodology that supports the findings of this study are available in the supplementary material.